\documentclass[12pt]{article}  

\usepackage{a41}
\usepackage{cite}
\usepackage{epsfig}       
\usepackage{graphicx}
\usepackage{amsmath}
\usepackage{amssymb}     
\usepackage{mdwlist}
\usepackage{color}

\usepackage[rflt]{floatflt}
\usepackage{float}
\usepackage{slashed}

\usepackage{array}

\usepackage[english]{babel}
\usepackage[latin1]{inputenc}
\usepackage[T1]{fontenc}
\usepackage{ae}

\usepackage{url}

\usepackage{amsmath, amsthm, amssymb}
\newtheorem{thm}{Theorem}[section]

\newtheorem{definition}[thm]{Definition}

\newcommand{\lsim}{\raisebox{-0.07cm   }
{$\, \stackrel{<}{{\scriptstyle\sim}}\, $}}

\newcommand{\GeV}{\rm GeV}

\usepackage{rotating}

\usepackage{graphicx}

\newcounter{mmacnt}
\def\restartmma{\setcounter{mmacnt}{0}}
\restartmma \catcode`|=\active
\def|#1|{\mathrm{#1}}
\catcode`|=12
\newenvironment{mma}{
 \par\smallskip
 \catcode`|=\active
 \parskip=0pt\parindent=0pt 
 \small
 \def\In##1\\{%
   \def\linebreak{\hfill\break\null\qquad}%
   \refstepcounter{mmacnt}
   \hangindent=2.5em\hangafter=0
   \leavevmode
   \llap{\tiny\sffamily In[\arabic{mmacnt}]:=\kern.5em}%
   \mathversion{bold}
\footnotesize$\displaystyle##1$\normalsize
   \mathversion{normal}\par
 }%
 \def\Print##1\\{%
   \def\linebreak{\hfill\break}%
   \hangindent=2.5em\hangafter=0
   \leavevmode ##1\par}%
 \def\Out##1\\{%
   \def\linebreak{$\hfill\break\null\hfill$}%
   \kern\abovedisplayskip\par
   \hangindent=2.5em\hangafter=0
   \leavevmode
   \llap{\tiny\sffamily Out[\arabic{mmacnt}]=\kern.5em}
   \footnotesize$\displaystyle##1$\normalsize\hfill\null\par
   \kern\belowdisplayskip
 }%
 \def\Warning##1##2\\{%
   \def\linebreak{\hfill\break}%
   \hangindent=2.5em\hangafter=0
   \leavevmode
   {\scriptsize##1 : ##2}\par}%
}{%
 \par\smallskip
}

\usepackage{color}

\newenvironment{fshaded}{%
\MakeFramed {\FrameRestore}
}%
{\endMakeFramed}

\allowdisplaybreaks[4]

\begin{document}
\setlength{\baselineskip}{0.515cm}
\sloppy
\thispagestyle{empty}
\begin{flushleft}
DESY 15-198
\hfill {\tt arXiv:1511.00229 [hep-ph]}
\\
DO--TH 15/15\\
October  2015
\end{flushleft}

\mbox{}
\vspace*{2cm}
\begin{center}

{\Large\bf A Kinematic Condition on Intrinsic Charm}

\vspace{2cm}
{\large
Johannes~Bl\"umlein}

\vspace{1cm}
\normalsize
{\it  Deutsches Elektronen--Synchrotron, DESY,}\\
{\it  Platanenallee 6, D-15738 Zeuthen, Germany}
\\


\end{center}
\normalsize

\vspace*{\fill}
\begin{abstract}
\noindent
We derive a kinematic condition on the resolution of intrinsic charm and discuss phenomenological consequences.
\end{abstract}

\vspace*{\fill}
\noindent
\newpage

\vspace*{1mm}
\noindent
Intrinsic charm in nucleons has been proposed as a phenomenon, which can be described in the light-cone
wave function formalism \cite{Brodsky:1980pb} using old fashioned perturbation theory \cite{Weinberg:1966jm}.
It is characterized by a Fock state
\begin{eqnarray}
|q_1\, q_2\, q_3\, Q_4\, Q_5 \rangle,
\label{eq:STATE}
\end{eqnarray}
with massless quarks $q_i$ and heavy quarks $Q_j$ of mass $M_Q$.\footnote{One-loop radiative corrections 
were 
calculated in \cite{Hoffmann:1983ah}}. The emergence of this state can be viewed as a definite 
quantum fluctuation in front of a general hadronic background, which can be resolved in deep-inelastic 
lepton-nucleon scattering. 

Extrinsic heavy flavor contributions \cite{Witten:1975bh,HQ}, on the other hand, are due to {\it factorized} single 
massless parton induced processes, exciting the heavy quark contributions. For neutral current interactions the process 
results from vector boson-gluon fusion \cite{Witten:1975bh} and appears in first order in the strong coupling 
constant 
$\alpha_s$ at the quantum level.

Both processes are distinct and  of very different nature. As has been shown in Ref.~\cite{Brodsky:1980pb} 
the intrinsic charm contributions are situated at larger values of  $x$, while major 
contributions of the extrinsic charm appear at low values of $x$. While the intrinsic charm 
contribution appears in the scaling limit already, extrinsic charm contributes on the quantum level only. 

In the following we  derive the condition under which intrinsic charm is unambiguously visible in deep-inelastic scattering. 
We follow Drell and Yan, Ref.~\cite{Drell:1970yt}, and compare the lifetime, $\tau_{\rm life}$, of the intrinsic 
charm state 
with the interaction time in the deep-inelastic process, $\tau_{\rm int}$, demanding
\begin{eqnarray}
\frac{\tau_{\rm life}}
     {\tau_{\rm int}} \gg 1
\label{eq:COND}
\end{eqnarray}
as a necessary criterion for the observation of the phenomenon. Eq.~(\ref{eq:COND}) delivered a clear condition on the applicability 
of the (massless) parton model singling out the corresponding ranges in $x$ and $Q^2$. Here the major requests are
that the virtuality $Q^2$ of the process is much larger than any transverse momentum squared in the hadronic wave-function, 
$Q^2 \gg k_{\perp,i}^2$, and the Bjorken variable $x$ 
shall neither get close to 1 nor take too small values, \cite{Drell:1970yt}.
Usually, in the excluded regions other contributions, like higher twist terms are present and/or there is a  need of novel 
small-$x$ resummations, which are both of comparable or even of larger size than the terms computed. In the following 
we will apply Eq.~(\ref{eq:COND}) to the case of the state (\ref{eq:STATE}).

In an infinite momentum frame we may express the momentum transfer by the electro-weak boson probing the nucleon, $q$, as 
follows \cite{Drell:1970yt} 
\begin{eqnarray}
q = (q_0; q_3, q_\perp),~~q_0 = \frac{2 m_p \nu +q^2}{4 P},~~q_3 = -\frac{2 m_p \nu  -q^2}{4 P}, 
\end{eqnarray}
where $q^2 = -Q^2$, $m_p$ the proton mass, $\nu$ the energy transfer to the nucleon in the proton rest 
frame,  and $P$ is the large (`infinite') momentum.

The interaction time  $\tau_{\rm int}$ is given by
\begin{eqnarray}
\tau_{\rm int} = \frac{1}{q_0} = \frac{4 P}{2 m_p \nu +q^2} = \frac{4 P x}{Q^2(1-x)}~. 
\label{eq:INT}
\end{eqnarray}
Here $x$ denotes the momentum fraction of the struck quark. Likewise, we obtain for the lifetime 
of the intrinsic charm state 
\begin{eqnarray}
\tau_{\rm life} = \frac{1}{\sum_i E_i - E} = \left. \frac{2 P}{\left(\sum_{i=1}^5 \frac{M_i^2 
+k_{\perp,i}^2}{x_i}\right) - m_p^2} \right|_{\sum_j x_j =1},
\end{eqnarray}
with $E_i = \sqrt{x_i^2 P^2 + M_i^2 +k_{\perp,i}^2}$ the energies of the partons in the state and $E$ the total 
energy, applying the infinite momentum representation, consistently neglecting sub-leading terms $\sim 1/P$ 
in the large momentum. $M_i$ denotes the mass of the $i$th quark, $k_{\perp,i}$ its transverse 
momentum, and $x_i$ its momentum fraction. Deriving $\tau_{\rm life}$  for intrinsic charm, we consider three 
massless valence quarks and the heavy quark-antiquark pair in the Fock state. We set the  masses of the three 
light valence quarks to zero  and neglect the effect of transverse momenta, as in the 
derivation of Eq.~(8) \cite{Brodsky:1980pb},  but retain the term $m_p^2/M_Q^2$ here. One obtains 
\begin{eqnarray}
\tau_{\rm life}(x) &=& \left. \frac{2 P}{M_Q^2} \int_0^{1-x_5} dx_4 \int_0^{1-x_4-x_5} dx_3 
\frac{1-x_3-x_4-x_5}{\frac{1}{x_4} + \frac{1}{x_5} - \frac{m_p^2}{M_Q^2}}\right|_{x_5 = x}
\nonumber\\
&=& \frac{P x}{6 M_Q^2 (1-c x)^4} 
\Biggl\{
(1-x) (1-c x)
   [ 2 +
x [5 - x - c (1-x) [4 + x (5 - 2 c (1-x))]]
]
\nonumber\\ &&
+6 x (1 - c x (1-x))^2 \ln
   \left[\frac{x}{1- c x (1-x)}\right]\Biggr\},
\label{eq:TLIV}
\end{eqnarray}
with $c = m_p^2/M_Q^2$. The integrals in (\ref{eq:TLIV}) are the same as used to derive the probability distribution
$P(x)$ in \cite{Brodsky:1980pb} and Eq.~(\ref{eq:PP}), however, the energy denominator appears in the first power.
\begin{figure}[H]
\centering
\includegraphics[width=0.45\textwidth]{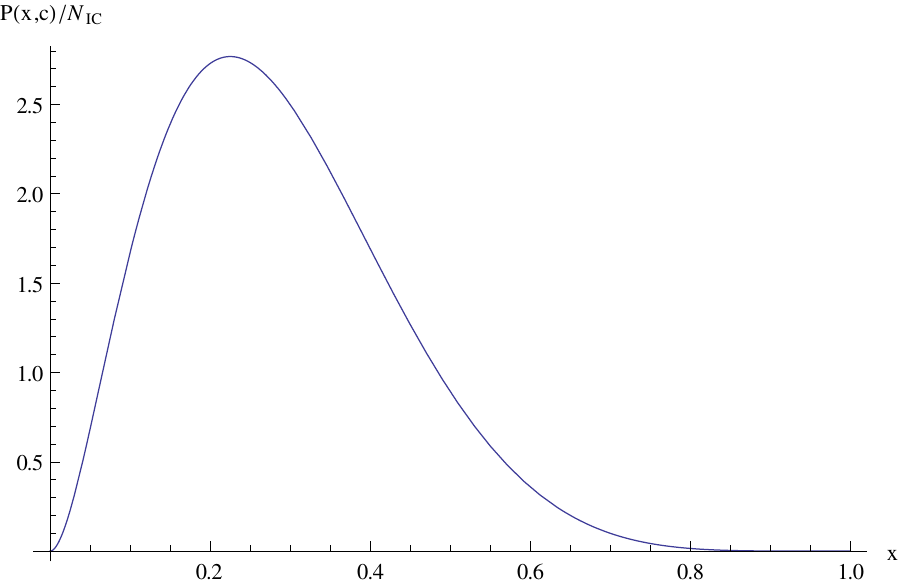}  
\includegraphics[width=0.45\textwidth]{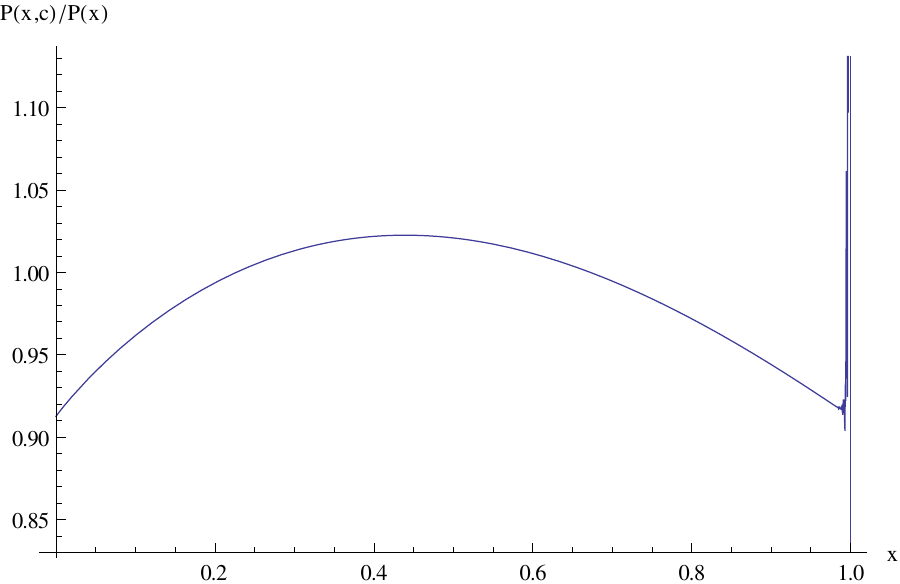}  
\caption[]{\sf \small 
Left panel: Normalized intrinsic charm distribution with finite $m_p$; Right panel: ratio of the probability distribution 
for intrinsic charm including the effect of the proton mass to the case $m_p/M_Q \rightarrow 0$. The parameter $c$ is given 
by $c \simeq 0.348$ and $M_Q = 1.59~\GeV$ in the pole mass scheme \cite{Alekhin:2012vu}.~\footnotemark} 
\label{fig:1}
\end{figure}
\footnotetext{Lower values of $m_c \sim 
1.3$~GeV 
used e.g. in \cite{Dulat:2013hea} at NLO are fully compatible with the NNLO value applied in the present study, 
cf.~\cite{Alekhin:2012vu}.
}

One may estimate also a lifetime for extrinsic $c\bar{c}$-production, if viewed as Fock state. Due to 
the factorized production, one considers the state  $|c \bar{c} X\rangle$, with $X$ the hadronic remainder of 
momentum fraction $x_1$, which yields
\begin{eqnarray}
\tau_{\rm life}^{\rm ext}(x) &=& \left. \frac{2 P}{M_Q^2} \int_0^1 dx_4 \int_0^1 dx_1 \delta(1 - x_1 - x_4 -x_5)
\frac{1}{\frac{1}{x_4} +\frac{1}{x_5} - c} \right|_{x_5 = x}
\nonumber\\
&=& \frac{2 P x}{M_Q^2 (1-c x)^2} \left\{(1-x)(1 - cx) + x \ln\left[\frac{x}{1- c x(1-x)}\right]\right\}~.
\label{eq:TLIVex}
\end{eqnarray}

The lowest order probability distribution for intrinsic charm, accounting for the nucleon mass effect, is 
given by
\begin{eqnarray}
P(x) &=& \left. N(c) \int_0^{1-x_5} dx_4 \int_0^{1-x_4-x_5} dx_3 \frac{1 - x_3 - x_4 - 
x_5}{\left(\frac{1}{x_4}+
\frac{1}{x_5} - c\right)^2} \right|_{x_5 = x}
\\
&=& \frac{N(c) x^2}{6 (1-c x)^5}
\Biggl\{
           (1-x) (1-c x) [1 + x[10 + x -c (1-x) (x (10-c (1-x))+2)]]
\nonumber\\ &&
+ 6 x [1+ x (1-c (1-x))] [1-c (1-x) x] 
[\ln(x)-\ln[1-c (1-x) x]]\Biggr\},
\label{eq:PP}
\end{eqnarray}
with $N(c)$ determined such that $\int_0^1 dx P(x) = N_{\rm IC}$, the integral fraction of 
intrinsic charm. Here we retained the effect of the proton mass, which was neglected in 
\cite{Brodsky:1980pb},  and illustrate the distribution in Figure~1a. One obtains a modification of the 
intrinsic charm distribution due to the finite nucleon mass effect of up to 10\%, as shown in
see Figure~1b. Setting $c \rightarrow 0$ leads to the previous result \cite{Brodsky:1980pb}
\begin{eqnarray}
P(x) &=& 600 N_{\rm IC} x^2 \Biggl[(1-x)(x^2+10 x +1) +6 x (x+1) \ln(x) \Biggr]~.
\end{eqnarray}

The ratio $\rho(x)= \tau_{\rm life}/\tau_{\rm int}$ is Lorentz-invariant and has to be larger than a suitable 
bound of $O(5 ... 10)$. The size of this value is fixed using standard requests applied also for the parameter setting
in experimental pulse resolution techniques e.g. in particle detectors. Here $\rho = 5$ would refer to a failure rate of 
20\% and $\rho = 10$ of 10\%\footnote{I would like to thank Dr. J. Bernhard from the Compass experiment for a corresponding 
remark.}.
\begin{figure}[H]
\centering
\includegraphics[width=0.5\textwidth]{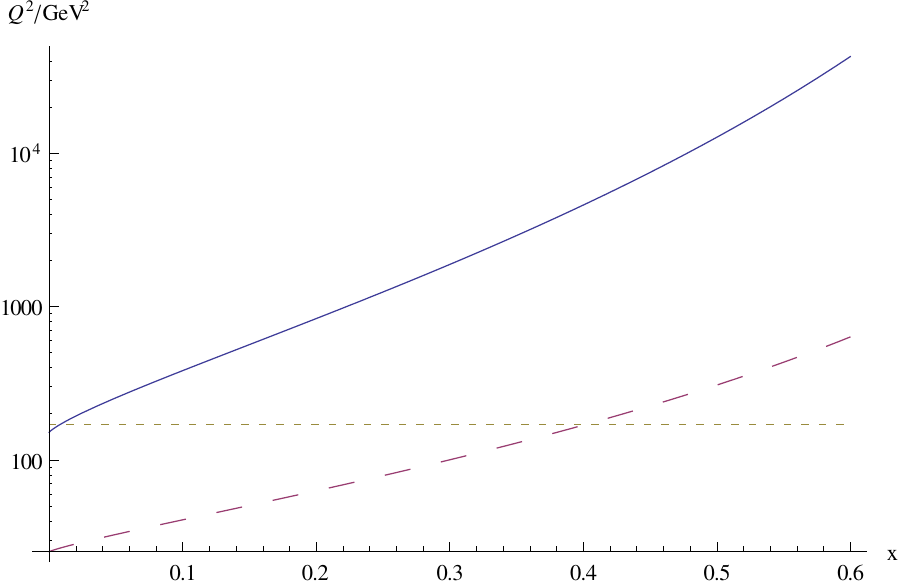}  
\caption{\sf \small Full line: lower boundary in $Q^2/\GeV^2$ for which 
$\rho(x) \geq 5$ as a function of 
$x$. Long dashed line: lower boundary for extrinsic charm production 
resulting from Eq.~(\ref{eq:TLIVex}) also for $\rho(x) \geq 5$;
Short dashed line: highest $Q^2$-bin of the EMC experiment \cite{Aubert:1985fx}.
\label{fig:2}} 
\end{figure}

Using this condition we may determine the allowed $Q^2$-range as a function of 
$x$ in which the quantum fluctuation leading to intrinsic charm can be unambiguously resolved by 
deep-inelastic scattering. Eqs.~(\ref{eq:INT},\ref{eq:TLIV}) lead to the function 
\begin{eqnarray}
\rho(x) \equiv \frac{\tau_{\rm life}(x)}{\tau_{\rm int}(x)} \geq \frac{Q^2}{12 M_Q^2} \simeq  
\frac{Q^2}{30.34~\GeV^2}.
\end{eqnarray}
The function $\rho(x)$  rises with growing values of $x$, i.e. a minimal bound of $Q^2 > 151~\GeV^2$ is obtained, 
demanding the 
ratio to be $\rho(x) \geq 5$. We show the corresponding $x$-dependence in Figure~2. We also show the boundary implied 
for extrinsic charm production. 

Let us consider the kinematics of the EMC experiment at CERN \cite{Aubert:1985fx}, which probably was the first measuring charm 
final states of a larger amount in deep-inelastic scattering
\cite{Aubert:1982tt}. These data have frequently been analyzed also searching for intrinsic charm. The highest $Q^2$ bin is centered 
at $Q^2 \simeq 170~\GeV^2$. The kinematic range  allowed for a clear intrinsic charm signal demanding $\rho(x) \geq 5$ is 
obtained as $x \lsim 0.01$, far below the peak-region at $x \simeq 0.22$ of the predicted distribution. On the other 
hand, the bound resulting for extrinsic charm production, cf. Eq.~(\ref{eq:TLIVex}), covers a  wider range, also of 
the kinematic region probed by the EMC experiment. Note that, furthermore, the accessible range in $x$ is 
strongly correlated to the 
probed region in $Q^2$ in deep-inelastic scattering experiments. This has to be taken into account interpreting low energy data as 
those of the EMC experiment in terms of intrinsic charm effects. Several phenomenological analyses have been carried out to 
search for intrinsic charm, cf. e.g. Refs.~\cite{Harris:1995jx}. Other analyses came to very similar conclusions of a possible 
integral fraction of $N_{\rm IC}$ in the range of up to $O(1..3\%)$. In all these analyses the life-time 
constraint (\ref{eq:COND}) has not been considered.

Discoveries need clean conditions. The $Q^2$ bound illustrated by Figure~2 points to a much more fortunate situation to search 
for intrinsic charm effects opening up at high energy colliders if compared to fixed target experiments, such as at HERA or 
within future projects like the EIC \cite{EIC} and LHeC \cite{AbelleiraFernandez:2012cc}, also operating at high luminosity. 
Condition (\ref{eq:COND}) is more easily fulfilled there because of the much wider kinematic range. As a consequence, 
intrinsic charm can be searched for in a dedicated way only at high energies.

\vspace{5mm}
\noindent
{\bf Acknowledgment.}~
I would like to thank H.~Fritzsch and G.~Branco for organizing a nice conference on High Energy 
Physics in the beautiful Algarve, where this note has been worked out. Conversations with S.~Brodsky 
and M.~Klein are gratefully acknowledged. This work was supported in part by the European Commission 
through contract PITN-GA-2012-316704 ({HIGGSTOOLS}).

{\small

}
\end{document}